\documentclass[sigconf,10pt]{acmart}
\usepackage{amsmath,amsfonts}
\usepackage{url}
\usepackage{verbatim}
\usepackage{graphicx}
\usepackage{subfigure}
\usepackage{caption}
\usepackage{graphicx}
\usepackage{algorithm}
\usepackage{algpseudocode}
\usepackage{textcomp}
\usepackage{multirow}
\usepackage{url}
\usepackage{color}
\usepackage{pifont}
\usepackage{xcolor}
\usepackage{tabularx}
\usepackage{pifont}
\usepackage{mdwlist}
\usepackage{booktabs}
\usepackage{makecell}
\usepackage{booktabs}
\usepackage{hyperref}

\newcounter{glossy_enum}
\makecompactlist{list*}{list}
\newenvironment{glossy_enumerate}
{\begin{list*}[{\arabic{glossy_enum})}{\usecounter{glossy_enum} \topsep=0.2em \leftmargin=1.4em \itemindent=-0.0em}]}
{\end{list*}\vspace*{0em}}

\newcommand{\fakepar}[1]{\vspace{.5mm}\noindent\textbf{#1.}}

\AtBeginDocument{%
  \providecommand\BibTeX{{%
    \normalfont B\kern-0.5em{\scshape i\kern-0.25em b}\kern-0.8em\TeX}}}

\setcopyright{rightsretained}

\begin{document}
\title{SparOA: Sparse and Operator-aware Hybrid Scheduling for Edge DNN Inference}

\author{Ziyang Zhang}
\affiliation{%
  \institution{Politecnico di Milano}
  \country{Milan, Italy}}
\email{ziyang.zhang@polimi.it}

\author{Jie Liu}
\affiliation{%
  \institution{Harbin Institute of Technology}
  \country{Shenzhen, China}}
\email{jieliu@hit.edu.cn}

\author{Luca Mottola}
\affiliation{%
  \institution{Politecnico di Milano}
  \country{Milan, Italy}}
\email{luca.mottola@polimi.it}


\renewcommand{\shortauthors}{xxx et al.}

\begin{abstract}
The resource demands of deep neural network (DNN) models introduce significant performance challenges, especially when deployed on resource-constrained edge devices. Existing solutions like model compression often sacrifice accuracy, while specialized hardware remains costly and inflexible. Hybrid inference methods, however, typically overlook how operator characteristics impact performance.
In this work, we present SparOA, a CPU-GPU hybrid inference framework, which leverages both sparsity and computational intensity to optimize operator scheduling.
SparOA embraces aforementioned challenges through three key components: (1) a threshold predictor that accurately determines optimal sparsity and computational intensity thresholds; (2) a reinforcement learning-based scheduler that dynamically optimizes resource allocation based on real-time hardware states; and (3) a hybrid inference engine that enhances efficiency through asynchronous execution and batch size optimization.
Extensive results show that SparOA achieves an average speedup of $1.22\times$$\sim$$1.31\times$ compared to all baselines, and outperforms the CPU-Only by up to 50.7$\times$.
Also, SparOA achieves optimal energy-per-inference, consuming 7\%$\sim$16\% less energy than the SOTA co-execution baseline.
\end{abstract}

\maketitle

\section{Introduction}\label{sec:introduction}
An increasing number of intelligent applications rely on deep neural network (DNN) models for real-time inference at the network edge~\cite{shi2016edge}. These applications span various fields, including but not limited to autonomous driving~\cite{tang2023torchsparse++}, industrial automation~\cite{fang2023joint}, and medical diagnostics~\cite{gongye2020new}. However, edge devices often have limited computational power, storage capacity, and energy supply, making it highly challenging to deploy and run complex DNN models on these devices. 

To address this challenge, software and hardware optimization methods exist in the literature. 
For instance, model compression~\cite{han2015deep} reduces model size and computational complexity through pruning, quantization, and knowledge distillation, making the models more suitable for deployment on resource-constrained devices. However, such methods often come at the cost of reduced inference accuracy and may fail to meet real-time requirements in certain scenarios. 
Leveraging specialized hardware, such as TPUs and FPGAs, may accelerate the inference process. While this hardware  can significantly enhance performance for specific types of tasks, their high cost and limited flexibility make them difficult to gain wide adoption for general-purpose edge devices.

\fakepar{Hybrid inference} Traditional solutions targeting efficient DNN inference also typically focus on a single hardware component, such as the CPU or GPU. 
However, these techniques often miss out on the potential gains due to employing multiple components.
Hybrid inference~\cite{jeong2022band,ling2022blastnet,wei2023nn} techniques dynamically schedule operators within a DNN across different processors. For instance, CoDL~\cite{jia2022codl} dynamically schedules operator execution by evaluating its affinity for heterogeneous processors, thereby optimizing latency. 

Most existing hybrid inference methods overlook the impact of operator characteristics, such as sparsity and computational intensity.
Significant differences in the computational characteristics and resource requirements of each operator indeed exist. 
For instance, convolution and fully connected operators typically involve dense matrix multiplication or convolution operations, requiring substantial computational resources. 
In contrast, activation and normalization operators demand less computation. 

\fakepar{Key insight} We simultaneously consider \emph{sparsity} and \emph{computational intensity} as metrics to quantify the resource requirements of operators. 
First, sparsity is an important indicator of the distribution of computational load in DNN models. 
Operators with high sparsity imply that most activation values are zero, allowing these operations to be skipped to reduce computational overhead, yet without affecting the output. 
Computational intensity refers to the number of floating-point operations (FLOPs) required per unit of input data, serving as a measure of the computational complexity and resource demands. 
Operators with high computational intensity typically involve dense matrix multiplications or convolutions and require higher computational resources. 
Convolution and fully connected operators, due to their compute-intensive nature, are better suited for execution on a GPU, while normalization and activation operators are more efficiently processed on a CPU.

Existing work~\cite{fan2023sparse,song2024powerinfer} primarily focuses on optimizing inference latency based on sparsity while neglecting computational intensity. 
For instance, even if an operator exhibits high sparsity, scheduling it to run on a CPU when it also has high computational intensity may result in suboptimal performance. 
Conversely, parallel computing enabled by GPUs is better suited for such tasks. 
Existing methods also often employ static scheduling strategies, determining the allocation of operators beforehand. 
This method cannot adapt to dynamically changing hardware states and input data, leading to suboptimal performance. 
For instance, in some cases, a GPU may fail to efficiently handle tasks with high computational intensity due to insufficient memory, and the task should then  switch to the CPU for execution.

\fakepar{Contribution} We present SparOA, an adaptive operator scheduling strategy based on sparsity and computational intensity.
By accurately analyzing the sparsity and computational intensity characteristics of each operator, SparOA can automatically and dynamically optimize the allocation strategy of operators to heterogeneous processors. 

\begin{figure}[tb]
\large
\centerline{\includegraphics[width=1\linewidth]{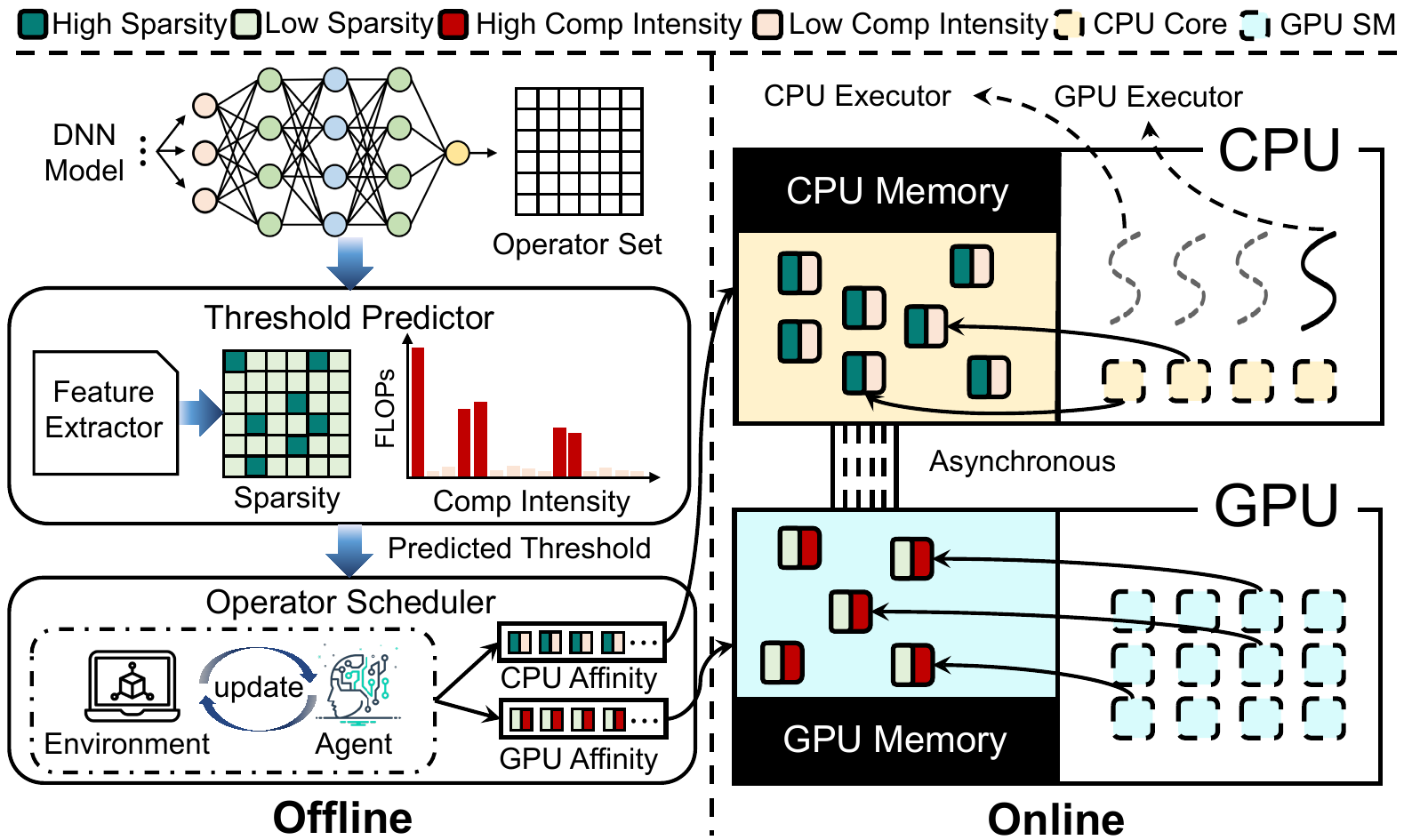}}
\caption{Overview of SparOA.}
\vspace{-0.3cm}
\label{fig:framework}
\end{figure}

Figure~\ref{fig:framework} shows an overview of SparOA, comprising both offline and online components.
The offline phase consists of two modules, i.e., the \emph{threshold predictor} and the \emph{operator scheduler}.
The light-weight, yet accurate threshold predictor estimates the optimal thresholds for sparsity and computational intensity of each DNN operator, guiding operator scheduling in the online phase. 
For instance, operators with high sparsity but low
computational intensity are prioritized for the CPU, while
operators with low sparsity and high computational intensity are preferentially assigned to the GPU.
Based on the predicted sparsity and computational intensity threshold, the operator scheduler optimizes the scheduling strategy for the input DNN model, using reinforcement learning. 
For the online phase, the \emph{hybrid inference} engine coordinates the synchronous execution of operators on heterogeneous processors according to the optimal scheduling strategy, while using asynchronous and dynamic batching to further improve resource utilization efficiency.

We implement SparOA using commercial off-the-shelf (COTS) edge devices, including NVIDIA Jetson AGX Orin and Orin Nano. 
We report in Sec.~\ref{evaluation} on experiments with popular DNN models including convolutional neural network (CNN)-based and attention-based architectures. 
The results show that SparOA consistently outperforms state-of-the-art (SOTA) methods, achieving an average $1.22\times$$\sim$$1.31\times$ speedup over SOTA compilers (e.g., TensorRT~\cite{NVIDIATensorRT}, TVM~\cite{chen2018tvm}) and co-execution frameworks (e.g., CoDL~\cite{jia2022codl}) , and up to 50.7$\times$ speedup over CPU-Only.

Overall, we make the following contributions:
\begin{glossy_enumerate}
\item
we preset a threshold predictor based on sparsity and computational intensity, which accurately predicts the resource requirements of operators;  
\item 
we conceive an operator scheduling strategy based on reinforcement learning, which dynamically optimizes the scheduling strategy according to real-time hardware status and input data;
\item 
we implement an efficient hybrid inference engine that significantly improves CPU-GPU hybrid inference efficiency through asynchronous execution and dynamic batching optimization.
\item 
the results show that SparOA achieves optimal latency and energy efficiency compared to the SOTA baselines.
\end{glossy_enumerate}

\section{Background and Motivation} \label{sec:background}
\subsection{Sparsity and Computational Intensity}\label{Sparsity}
\emph{Sparsity}~\cite{fan2023sparse,song2024powerinfer} refers to the proportion of zero elements in the activation values and is often used to measure the computational load distribution of a DNN model. High sparsity in an operator means that most of the activation values are zero, allowing these zero-value operations to be skipped. 

Evaluating sparsity entails capturing the activation value distribution characteristics of each DNN operator. For instance, the ReLU activation function typically produces high sparsity because negative values are directly set to zero. 
Operators with high sparsity, due to their reduced computational workload and lower data transfer latency, are more suitable for execution on the CPU. 
Conversely, operators with low sparsity require more computational resources and are therefore better suited for GPU processing. 

Sparsity may also be used for compressing and optimizing DNN models~\cite{fan2023sparse,zhu2023pockengine}. 
For instance, pruning operators with high sparsity may further reduce the computational complexity and storage requirements of a DNN. 
Further, sparsity analysis may help identify redundant operations within the DNN model, supporting operator fusion.

\emph{Computational intensity} measures the computational demand of each operator in a DNN, expressed in floating-point operations per second (FLOPs). 
Operators with high computational intensity may involve dense matrix multiplications, convolution operations, or complex nonlinear transformations. 
In DNN inference optimization, computational intensity analysis is essential.

For instance, a GPU with parallel computing capabilities is apt to handle tasks with high computational intensity, such as convolutions and fully connected operators. 
These operators may involve large amounts of matrix multiplications and additions, which can be efficiently completed on the GPU. 
In contrast, a CPU excels in single-threaded performance, making it more suitable for tasks with low computational intensity, such as normalization and activation functions. 

\begin{figure}[tb]
\large
\centerline{\includegraphics[width=0.85\linewidth]{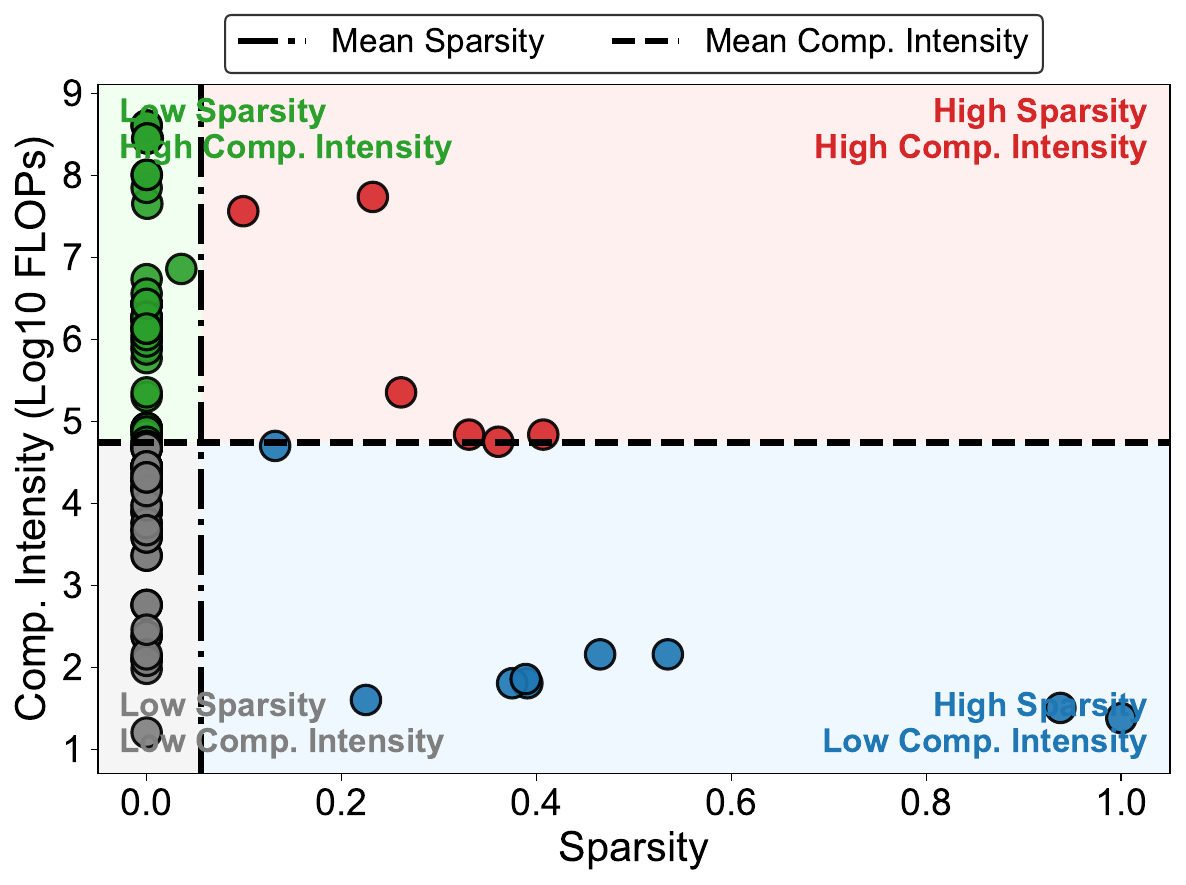}}
\vspace{-0.2cm}
\caption{The distribution of sparsity and computational intensity for each operator in MobileNetV3-small, measured on NVIDIA Jetson AGX Orin. We set the batch size to 1.}
\vspace{-0.3cm}
\label{fig:motivation}
\end{figure}

\subsection{Motivation}\label{motivation}
When scheduling operators, it is necessary to consider computational intensity to determine the hardware component for each task. 
Relying solely on sparsity for scheduling while ignoring the impact of computational intensity may lead to suboptimal performance. Certain operators, despite having high sparsity, may also exhibit high computational intensity, and incorrectly assigning them to the CPU for execution may result in increased latency and resource waste.

\fakepar{Independence of Sparsity and Computational Intensity}
Existing hybrid scheduling methods~\cite{song2024powerinfer,fan2023sparse} assume a strong negative correlation between sparsity and computational intensity.
However, our profiling reveals that these two metrics are, in fact, orthogonal dimensions. Figure~\ref{fig:motivation} illustrates the operator distribution of MobileNetV3-Small across the sparsity-intensity spectrum on the NVIDIA Jetson platform. The analysis uncovers four distinct quadrants:

\begin{itemize}
\item 
\emph{Quadrant II (High Sparsity \& High Intensity):} 
as shown in the top-right of Figure 2, certain operators (e.g., \emph{Conv2d} operator) exhibit significant sparsity ($>0.4$) due to accumulated \emph{ReLU} activations in prior layers. However, due to their large channel dimensions, their computational intensity remains extremely high ($>10^8$ FLOPs). Assigning these operators to the CPU solely based on their high sparsity would cause severe computational bottlenecks, as the CPU lacks the parallelism to handle the massive workload despite the zeros. 
\item 
\emph{Quadrant III (Low Sparsity \& Low Intensity):} 
conversely, the bottom-left quadrant contains operators like \emph{BatchNorm2d}. These operators are dense (sparsity $\approx 0$) but possess low arithmetic intensity. They are memory-bound rather than compute-bound. Scheduling them to the GPU based solely on their density would incur unnecessary kernel launch and data transfer overheads that outweigh the execution benefits.
\item 
\emph{Quadrants I \& IV:} 
these represent the intuitive cases: dense, heavy operators in quadrant 2 ideal for GPU, and sparse, light operators in quadrant 4 ideal for CPU.
\end{itemize}

This distribution reveals that neither sparsity nor computational intensity alone is a sufficient proxy for CPU-GPU hybrid operator scheduling. We addresses this by treating them as joint state features, enabling the scheduler to distinguish these complex scenarios. For instance, advisably offloading high-sparsity/high-intensity operators to the GPU while keeping low-sparsity/low-intensity operators on the CPU to minimize end-to-end latency.

\section{Threshold Predictor} \label{sec:predictor}
Existing works usually rely on hand-designed rules to guide the scheduling strategy of operators, which may simply assign all operators with sparsity above a fixed threshold to the CPU and other operators to the GPU. 
However, these methods show several shortcomings.
First, the fixed threshold is difficult to adapt to the diversity of different DNN models and input data.
Second, it ignores the impact of computational intensity on inference performance, which may lead to suboptimal scheduling decisions. 

\begin{figure}[tb]
\vspace{-0.5cm}
\large
\centerline{\includegraphics[width=0.85\linewidth]{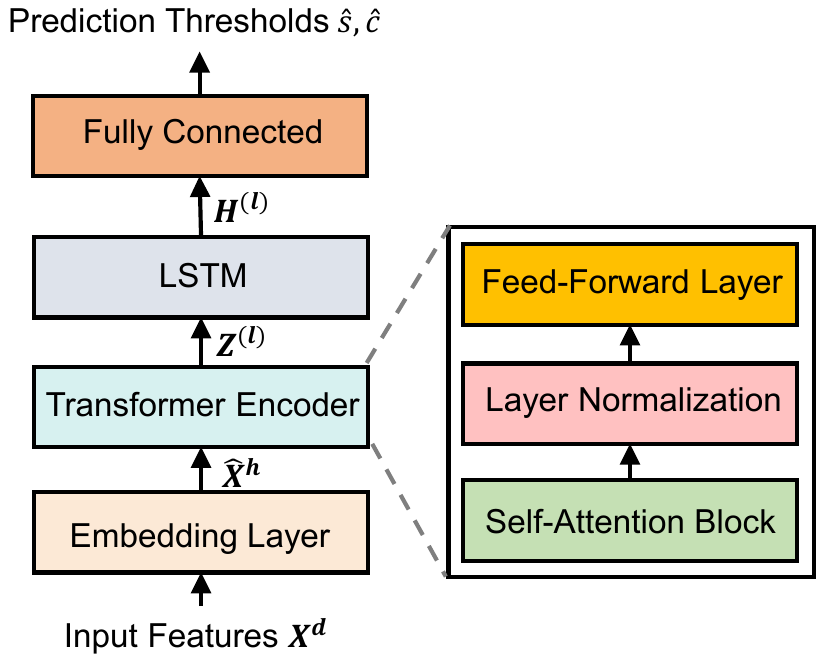}}
\vspace{-0.3cm}
\caption{The architecture of threshold predictor.}
\vspace{-0.5cm}
\label{fig:predictor}
\end{figure}

As shown in Figure~\ref{fig:predictor}, we leverage the advantages of the Transformer architecture in capturing long-range dependencies and contextual information, and design a Transformer-based threshold predictor to dynamically adjust the sparsity threshold and computational intensity threshold of each operator through automatic learning, and use it as the basis for scheduling decisions.

\subsection{Data Preprocessing and\\ Feature Extraction}
The input to the threshold predictor comes from statistical data of each operator in the DNN model, including sparsity, computational intensity and input features. 
The feature extraction process is as follows:
\begin{itemize}
\item \emph{Sparsity:} for each activation operator, we calculate sparsity by counting the proportion of non-zero elements. Let the output tensor of a certain layer be $O \in \mathbb{R}^{H\times W \times C}$, where $H,W,C$ represent height, width, and number of channels, respectively. Sparsity $\rho$ can be expressed as
\begin{equation}
\rho = 1 - \frac{c(O)}{n(O)}
\label{1}
\end{equation}
where $c(O)$ represents the number of non-zero elements in the tensor, and $n(O)$ represents the total number of elements in the tensor. 
\item 
\emph{Computational intensity:} for each compute-intensive operator, we estimate computational intensity by calculating the number of floating-point operations (FLOPs). 
For instance, the computational intensity of convolution operator $I$ can be expressed as:
\begin{equation}
I= K_h\cdot K_w\cdot C_{in}\cdot C_{out}\cdot H\cdot W
\label{2}
\end{equation}
where $K_h$, $K_w$ are the height and width of the convolution kernel, and $C_{in}$, $C_{out}$ are the number of input and output channels, respectively.
\end{itemize}

Instead, the \emph{input features} include batch size $B$, number of input channels $C_{in}$, input height $H$, and width $W$, extracted directly from the input tensor.

\subsection{Predictor Architecture}
This section details the architectural design of the threshold predictor. The primary motivation for this architecture is to capture both global dependencies between input features (sparsity, computational intensity, batch size, and tensor dimensions) and temporal correlation across sequential operators. Conventional methods, such as fixed threshold rules, fail to adapt to the diversity of DNN models and dynamic input data, resulting in low prediction accuracy and suboptimal resource utilization. To address this, our objective is to enhance prediction precision while supporting real-time inference optimization through a parameter-efficient yet compact neural network architecture. 

Our design combines three key components: a Transformer encoder to model long-range feature interactions, an LSTM module to capture sequential operator patterns, and a lightweight output layer for threshold prediction. This hybrid design enables the predictor to: \emph{i)} comprehensively analyze operator-processor affinity through multi-dimensional feature embedding, \emph{ii)} maintain temporal consistency in scheduling decisions for sequential DNN layers, and \emph{iii)} generate precise sparsity/computational intensity thresholds that directly optimize hybrid execution performance metrics. 

The design of each component is described in detail below.
\begin{itemize}
\item
\emph{Embedding layer:} maps the input vector $X=[\rho,I,B,C_{in}, \\ H,W], X\in \mathbb{R}^d$ into a high-dimensional space to capture complex relationships between features. The output dimension of the embedding layer is $h$ (hidden layer dimension).
\item 
\emph{Transformer encoder:} the Transformer encoder consists of multiple self-attention mechanisms and feed-forward networks, used to capture global dependencies between features. The output of each encoder can be represented as:
\begin{equation}
Z^{(l)}=\operatorname{FFN}(\operatorname{LN}(X^{(l-1)}+\operatorname{MHSA}^{(l-1)})))
\label{3}
\end{equation}
where $l$ represents the number of Transformer encoder layers, $\operatorname{MHSA}$ represents multi-head self-attention, $\operatorname{FFN}$ represents feed-forward network, and $\operatorname{LN}$ represents normalization.
\item 
\emph{LSTM:} to capture temporal dependencies in sequential features, the output of the Transformer is fed into a bidirectional $\operatorname{LSTM}$ (Long Short-Term Memory~\cite{hochreiter1997long}), the output can be represented as:
\begin{equation}
H^{(l)}_t=\operatorname{LSTM}(Z^{(l)}_t)
\label{4}
\end{equation}
where $Z^{(l)}_t$ is the output sequence of the Transformer at time step $t$, and $H^{(l)}_t$ is the hidden state of the $\operatorname{LSTM}$ at time step $t$.
\item 
\emph{Fully connected:} the output of the $\operatorname{LSTM}$ at the last time step $T$, finally, is mapped to sparsity and computational intensity thresholds through a fully connected layer. The output can be represented as:
\begin{equation}
(\hat{s}_t,\hat{c}_t)=\sigma(\omega\cdot H^{(l)}_T +b)
\label{5}
\end{equation}
where $\hat{s}_t$ and $\hat{c}_t$ represent the predicted sparsity and computational intensity thresholds, respectively. 
$\omega$ and $b$ are weight and bias parameters, and $\sigma$ is the activation function (e.g., Sigmoid or ReLU).
\end{itemize}

\subsection{Training}
The training of the threshold predictor adopts a supervised learning method. 
Let the training dataset be $\{(X_i, y_i) \}^N_{i=1}$, where $X_i$ is the input vector, and $y_i=(s_i, c_i)$ are the actual sparsity and computational intensity thresholds. 
The loss function is defined as
\begin{equation}
\mathcal{L} = \frac{1}{N} \sum_{i=1}^{N} \left( | s_i - \hat{s}_i |^2 + | c_i - \hat{c}_i |^2 \right)
\label{6}
\end{equation}
where $N$ is the number of samples, $s_i$ and $c_i$ represent the true sparsity and computational intensity thresholds.

\fakepar{Ground Truth Acquisition}
To obtain the ground truth labels $y_i = (s_i, c_i)$ for supervised training, we conduct a one-time, offline exhaustive search on the target hardware. For each operator type in our dataset, we measure its execution latency across a comprehensive grid of sparsity levels and input sizes (which correlate to computational intensity) on both the CPU and GPU. The true optimal thresholds $(s_i, c_i)$ are identified as the boundary points where the optimal execution device switches (e.g., from CPU to GPU) to minimize latency. 
Summarily, we collected approximately 2,000 samples from the five DNN models listed in Table~\ref{tbl:model} on both Jetson Orin Nano and AGX Orin.

\section{Operator Scheduler} \label{Scheduling}
The primary goal of the learning-based operator scheduler in SparOA is to minimize the total inference time by intelligently assigning operators to heterogeneous hardware components. This involves considering multiple factors, including sparsity, computational intensity, memory constraints, and overheads of switching between different processors. 
We employ the Soft Actor-Critic (SAC) algorithm, a state-of-the-art RL method, to learn the optimal policy for operator scheduling. 
The algorithm is well-suited for dynamic, high-dimensional search spaces, making it applicable to the complex operator scheduling decisions at stake.

\fakepar{Learning vs. Rules}
Note that SparOA does not rely on fixed heuristic rules (e.g., always assign high-intensity operators to GPU). Instead, sparsity ($\rho$) and computational intensity ($I$) serve merely as state features input to the SAC agent. The agent autonomously learns the complex, non-linear mapping between these features and the optimal device assignment through trial-and-error, guided by a reward function that penalizes latency. Consequently, the agent retains the flexibility to make counter-intuitive but optimal decisions—for instance, scheduling a high-intensity operator on the CPU if the GPU memory is contended or if the data transfer overhead outweighs the acceleration benefit. This learning-based approach allows SparOA to adapt to specific operator behaviors (e.g., sparse vs. dense kernels) beyond simple intensity metrics.

\subsection{Modeling}
We model the operator scheduling problem as a Markov decision process (MDP). 
We defined next its components including the state space, the action space, as well as the reward and transition probabilities.

\fakepar{State space} We model the current state of reasoning process and used that to provide a decision-making basis for agents in RL. The state space includes the following features:
\begin{itemize}
\item \emph{Sparsity:} 
operators with high sparsity can usually reduce the amount of computation by skipping zero-value operations. 
\item
\emph{Computational intensity:} indicates the number of FLOPs required per unit of input data.  
\item 
\emph{Input size:} the  size of the input tensor for the current operator.
\item 
\emph{Output size:} the size of the output tensor for the current operator, which is used to evaluate the computational requirements of subsequent operators.
\item 
\emph{Hardware state:} includes GPU memory usage, CPU load level, and switching overhead between the heterogeneous processors.
\end{itemize}
Therefore, the state space $\mathcal{S}$ in RL is defined as 
\begin{equation}
\mathcal{S}=\{\rho, I, N_{in}, N_{out}, M_{gpu}, M_{cpu}, O_{switch}\}
\label{state}
\end{equation}
where $\rho$ is the sparsity, $I$ is the computational intensity, $N_{in}$ and $N_{out}$ are the input and output sizes respectively. $M_{gpu}$ is the GPU memory usage, $M_{cpu}$ is the CPU load level, and $O_{switch}$ is the heterogeneous processor switching overhead.

\fakepar{Action space} The action space in SparOA is a continuous range that represents the resource allocation ratio between the CPU and the GPU. Specifically, the action $\mathcal{A} \in [0,1]$ represents the proportion allocated to the GPU, whereas the remainder is allocated to the CPU. For instance, $\mathcal{A}=1$ means full use of the GPU and $\mathcal{A}=0$ means full use of the CPU.
The action space can be formalized as follows:
\begin{equation}
a \in \mathcal{A}, \mathcal{A} \in [0,1]
\label{action}
\end{equation}

\fakepar{Reward} We formulate the as an evaluation of the quality of the agent's behavior in RL. In SparOA, the reward takes into account the following factors:
\begin{itemize}
\item 
\emph{Inference latency:} encourages reducing inference time. 
\item
\emph{Memory usage:} avoids exceeding memory limits. 
\item 
\emph{Switching overhead:} reduces processor switching.
\end{itemize}
The reward is formalized as 
\begin{equation}
r = -(\lambda_1\cdot L+ \lambda_2\cdot (M_{gpu}+ M_{cpu}) + \lambda_3\cdot O_{switch})
\label{reward}
\end{equation}
where $L$ is the inference latency and $\lambda$ is the weight coefficient used to balance the importance of different goals.

\fakepar{Transition probabilities}
It mathematically formalizes how the environment's state evolves when an action is taken. Specifically, it defines the probability distribution $ P(s'|s,a)$, representing the likelihood of transitioning to state $s'$ from state $s$ after applying action $a$. We derive this probability distribution by combining two key factors:  \begin{itemize}
\item 
\emph{DNN architectural constraints:} the dependencies between operators (e.g., layer execution order) and their inherent computational properties (e.g., sparsity and computational intensity) impose deterministic transitions. 
\item
\emph{Hardware dynamic:} probabilistic variations arise from hardware-specific factors like GPU memory contention (e.g., when multiple operators compete for limited memory bandwidth) or CPU core availability (e.g., due to background processes). 
\end{itemize}

Consider a scenario where the current state $s$ includes a convolutional operator with sparsity $\rho=0.2$, computational intensity $I=1.8\times 10^{11} \,\text{FLOPs}$, and GPU memory usage $M_{\text{gpu}}=75\%$. The agent selects action $a=1$ (allocate to GPU). The transition to $s'$ is influenced by:  
1) The convolution's output tensor size $N_{\text{out}}$ updates the subsequent operator's input size.  
2) GPU memory usage increases by $\Delta M_{\text{gpu}}$, calculated from the operator's weight and activation storage requirements.  

\subsection{SAC-based Operator Scheduling}
SAC is an RL algorithm based on the maximum entropy framework, which achieves a balance between exploration and exploitation by introducing an entropy regularization term in the policy optimization process. 
The core idea is to maximize not only the expected return but also the entropy of the policy, thereby encouraging the agent to maintain a certain degree of randomness when making decisions. This feature makes it particularly suitable for solving complex continuous action space problems, such as tasks that dynamically allocate computing resources~\cite{sung2023decentralized}, that is, precisely our case. The key components of SAC include:
\begin{itemize}
\item \emph{Policy network:} $\pi_{\theta}(a|s)$ maps states $s$ to actions $a$. It is parameterized by $\theta$ and outputs a probability distribution over continuous actions. 
\item \emph{Q-networks:}  $Q_{\phi_1}$ and $Q_{\phi_2}$ estimate the expected return for taking action $a$ in state $s$. These two networks aim to mitigate overestimation, which can be updated using the Bellman equation: 
\begin{equation}
   Q(s, a) \gets r + \gamma \mathbb{E}_{s' \sim p, a' \sim \pi}[Q(s', a') - \alpha \log \pi(a'|s')]
\label{Q}
\end{equation}
where $ \gamma $ is the discount factor, and $ \alpha $ is the temperature parameter controlling the trade-off between exploration and exploitation.
\item \emph{Entropy regularization:} to maximize the expected return while maintaining high entropy in the policy:
\begin{equation}
J(\pi) = \mathbb{E}_{s \sim p, a \sim \pi}[Q(s, a)] + \alpha \mathcal{H}(\pi(\cdot|s))
\label{en}
\end{equation}
where $ \mathcal{H}(\pi(\cdot|s)) = -\mathbb{E}_{a_t \sim \pi}[\log \pi(a_t,s_t )]$ is the entropy of the policy.
\item 
\emph{Target networks:} to stabilize training, the parameters of the target networks are updated slowly using a moving average.
\begin{equation}
   \phi_i' \gets \tau \phi_i + (1 - \tau) \phi_i'
\label{target}
\end{equation}
where $ \tau $ is the smoothing coefficient.
\item 
\emph{Temperature parameter $\alpha$:} to dynamically adjust the strength of entropy regularization, SAC introduces a learnable temperature parameter $\alpha$, and the optimization objective becomes
\begin{equation}
J(\alpha) = \mathbb{E}_{a \sim \pi}[-\alpha \log \pi(a_t,s_t )+\bar{\mathcal{H}}]
\label{alpha}
\end{equation}
where $\bar{\mathcal{H}}$ is the target entropy, usually set to the negative action space dimension: $\bar{\mathcal{H}}=-dim(\mathcal{A})$.
\end{itemize}

Algorithm \ref{alg:sac_hybrid_inference} illustrates the operator scheduling optimization process. 
The algorithm operates in episodes, where it samples actions from the policy network to determine CPU-GPU resource ratios and executes computations accordingly (\emph{line 6$\sim$12}), either on both devices with weighted averaging (\emph{line 13}), solely on the CPU (\emph{line 15}), or solely on the GPU (\emph{line 17}). Transitions, including states, actions, rewards, and next states, are stored in the replay buffer (\emph{line 19}), while total inference latency is updated based on rewards (\emph{line 21}). After each episode, gradient updates improve the Q-functions and policy using Bellman backup, maximum entropy objectives, and dynamic temperature adjustment (\emph{line 23$\sim$30}).
By dynamically adjusting the temperature parameter $\alpha$, the algorithm maintains high exploration levels during the early stages of training and gradually converges to the optimal policy as training progresses.
\begin{algorithm}[t]
\footnotesize
\caption{SAC-based operator scheduling.}
\label{alg:sac_hybrid_inference}
\begin{algorithmic}[1]
\Require DNN model $ M $, input tensor $ T $, environment $ E $
\Ensure Operator Scheduling Strategy $ P $

\State Initialize policy network $ \pi_\theta(a | s) $, Q-networks $ Q_{\phi_1}(s, a), Q_{\phi_2}(s, a) $, target networks $ Q_{\phi_1}^{\text{target}}, Q_{\phi_2}^{\text{target}} $
\State Initialize replay buffer $ \mathcal{D} $, temperature parameter $ \alpha $
\State $ L \gets 0 $ \Comment{Initialize inference latency}

\For{$ \text{episode} = 1, 2, \dots $}
    \State $ S \gets E.\text{reset()} $
    \For{$ t = 1, 2, \dots $}
        \State Sample action $ \mathcal{A} \sim \pi_\theta(\cdot | \mathcal{S}) $ 
        \State Execute action $\mathcal{A}$ using SAC policy $\pi_\theta(\mathcal{A} | \mathcal{S})$
        \State $\xi = \text{map\_action\_to\_ratios}(\mathcal{A})$
        \If{$ \xi \in (0,1) $}
            \State $ P_{\text{cpu}} \gets \text{execute\_on\_cpu}(M, T) $
            \State$ P_{\text{gpu}} \gets \text{execute\_on\_gpu}(M, T) $
            \State $ P \gets \xi \cdot P_{\text{cpu}} + (1-\xi) \cdot P_{\text{gpu}}$
        \ElsIf{$ \xi = 0 $}
            \State $ P \gets \text{execute\_on\_cpu}(M, T) $
        \Else
            \State $ P \gets \text{execute\_on\_gpu}(M, T) $
        \EndIf
        \State $ \mathcal{D} \gets  (\mathcal{S}, \mathcal{A}, r, \mathcal{S}', \text{done}) $
        \State $\mathcal{S}', r, \text{done}, \_ \gets E.\text{step}(\mathcal{A})$
        \State $L \gets L - r$
    \EndFor
    \For{gradient step $g= 1, 2, \dots$}
         \State Sample mini-batch $ B = \{(s, a, r, s', \text{done})\} $ from $ \mathcal{D} $
        \State Compute target Q-values:
        \State Update $ Q_\phi(\mathcal{S}, \mathcal{A}) $ using Bellman backup
        \State Update $ \pi_\theta(\mathcal{A} | \mathcal{S}) $ via maximum entropy objective
        \State Update $ \alpha $ dynamically
        \State Perform soft updates for target networks
    \EndFor
\EndFor
\end{algorithmic}
\end{algorithm}

\section{Hybrid Inference Engine} \label{Engine}
The online hybrid inference engine coordinates the execution of operators on the CPU and GPU according to the scheduling policy generated by the operator scheduler. Traditional hybrid inference methods usually face challenges such as processor switching overhead and load imbalance. On the one hand, frequent data transfers between the CPU and GPU introduce significant latency. On the other hand, static resource allocation strategies, such as fixed batch sizes, cannot adapt to the dynamic characteristics of different operators and may cause memory overflows. 

We employ asynchronous executions and dynamic batching to further improve inference efficiency while minimizing data transfer overheads and performance losses caused by unbalanced resource allocation. 
Asynchronous execution parallelizes task scheduling through CUDA streams, reducing the overhead of device switching and data transmission. 
Dynamic batching optimization dynamically adjusts the batch size according to input data characteristics and hardware constraints to improve resource utilization.

\subsection{CPU-GPU Hybrid Execution}
\fakepar{Data transfer} To minimize the extra cost of collaboration, particularly the latency induced by CPU-GPU data movement and synchronization—we implement a highly optimized data transmission pipeline. We utilize pinned host memory (i.e., page-locked memory) rather than standard pageable memory for all CPU-resident tensors involved in co-execution. Pinned memory enables the DMA (i.e., direct memory access) controller to transfer data directly between system RAM and GPU VRAM without CPU intervention, significantly maximizing bandwidth utilization.

Furthermore, we leverage CUDA Streams to strictly enforce asynchronous execution. Specifically, we issue asynchronous memory copy commands using \texttt{cudaMemcpyAsync} in non-default streams. This allows the GPU to execute inference kernels for the current batch while simultaneously fetching the next batch's data from the CPU, or vice versa, effectively hiding the transmission overhead behind computation. We explicitly quantify this collaboration cost by profiling the duration of these memory operations, and evaluated it in Sec.~\ref{Overall}.

\fakepar{Result aggregation}
After an operator completes the execution on the CPU or GPU, we aggregate the results using the GPU or CPU executors (i.e., threads running on the CPU).
We employ a weighted average aggregation strategy with weights dynamically generated by the SAC-based scheduling algorithm.
We formalize the aggregation result as:
\begin{equation}
P = \xi \cdot P_{\text{cpu}} + (1-\xi) \cdot P_{\text{gpu}}
\label{Aggregation}
\end{equation}
where $P_{\text{cpu}}$ and $P_{\text{gpu}}$ are the outputs of the CPU and GPU respectively, and $\xi$ is the allocation ratios. 

The engine synchronizes the CUDA stream via \texttt{torch.cuda\allowbreak.synchronize} before aggregation to ensure the correctness of result aggregation.
Additionally, the engine caches the intermediate results of the CPU and GPU in shared memory for direct access by subsequent operations, thereby reducing the overhead of repeated calculations and data transfers.
Note that the operators preloaded in GPU are processed in situ, while the CPU calculates and transfers results for its operators to the GPU for aggregation~\cite{song2024powerinfer}.

\begin{figure}[tb]
\vspace{-0.2cm}
\large
\centerline{\includegraphics[width=\linewidth]{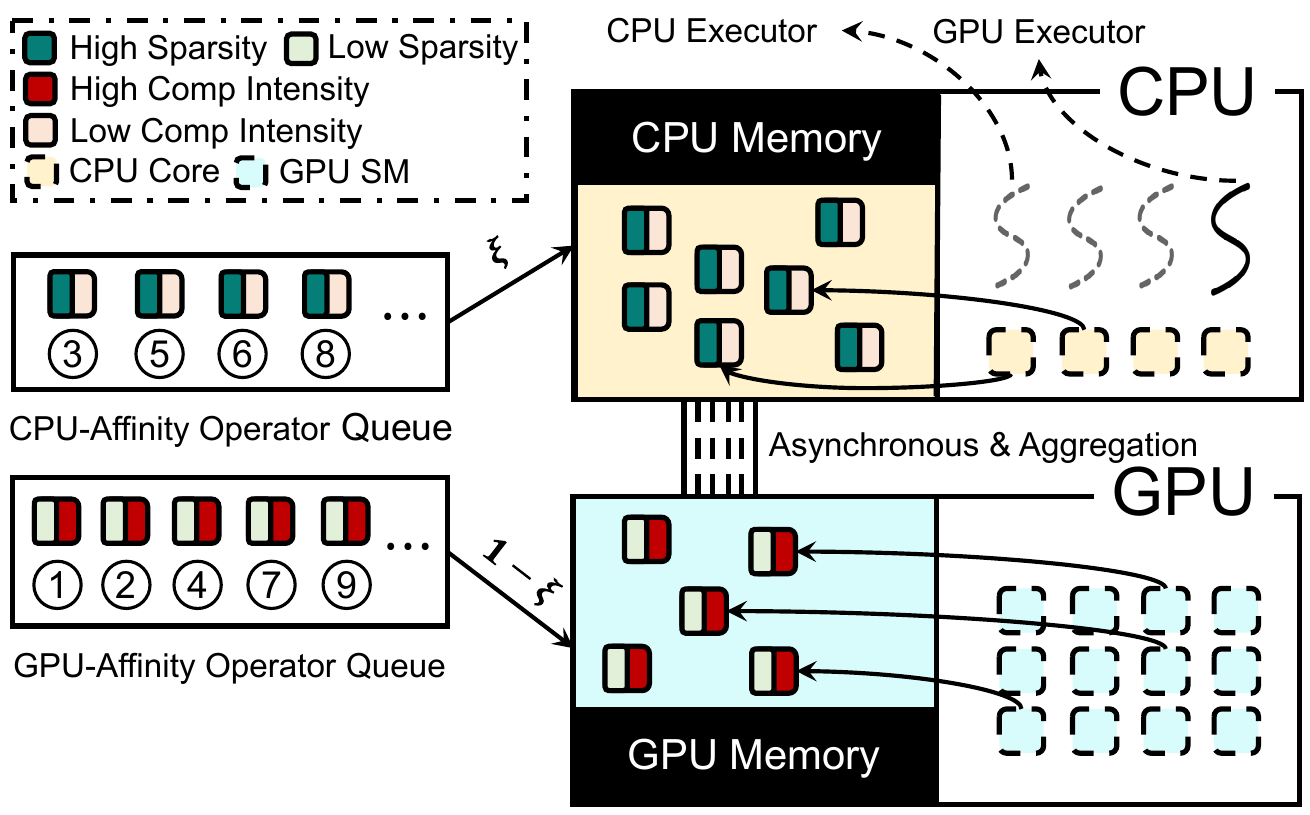}}
\vspace{-0.3cm}
\caption{Example showing how SparOA calculates different operators.}
\vspace{-0.5cm}
\label{fig:engine}
\end{figure}

Figure~\ref{fig:engine} illustrates an example of how SparOA co-executes operators across CPU and GPU. It classifies operators based on the proposed offline operator scheduler, assigning operators with high sparsity and low computational intensity (e.g., operators 3, 5, 6, 8) to CPU memory and others to GPU memory. 
Both the CPU and GPU concurrently process their assigned operators. 
The GPU computes operators 1, 2, 4, 7, and 8, leveraging its parallel processing capabilities, while the CPU simultaneously handles operators 3, 5, 6, 8. 
Upon completion of all operators on the CPU, its output is transferred to the GPU that performs the final aggregation step, combining the results from all activated operators using a weighted average strategy.

\subsection{Dynamic Batching}
Dynamic batching aims to dynamically adjust the batch size based on input data characteristics and hardware constraints, thereby improving throughput while retaining inference accuracy. 
First, we must ensure that the batch size does not exceed available memory and meets real-time constraints; second, we must adjust the batch size according to operator sparsity and computational intensity to balance computation and data transfer overhead.

To this end, we use a gradient descent-based dynamic batching optimization algorithm, combining hardware-aware and input data-driven partitioning strategies to dynamically adjust the batch size. As shown in Algorithm \ref{alg:dynamic_batch_size}, the gradient is calculated based on the inference latency and memory usage of the current batch size. The batch size is then gradually adjusted along the gradient direction until a (possibly local) optimal value is found (\emph{line 6}). 

During this process, the algorithm continuously checks whether the batch size satisfies hardware  and real-time constraints (\emph{line 7$\sim$14}). If the current batch size exceeds available memory, it is halved to avoid overflow. 
Additionally, the algorithm partitions the batch based on the sparsity and computational intensity characteristics of the input data. For instance, for highly sparse inputs, the batch size can be increased to leverage the parallel computing capabilities of the GPU; for inputs with high computational intensity, we reduce the batch size to shrink inference times.
\begin{algorithm}[t]
\footnotesize
\caption{Dynamic batching optimization.}
\label{alg:dynamic_batch_size}
\begin{algorithmic}[1]
\Require Initial batch size $ B_0 $, input tensor $ T $
\Ensure Optimized batch size $ B^* $

\State Initialize $ B \gets B_0 $
\State Evaluate initial latency $ L(B) $ and memory usage $ M(B) $
\State Set learning rate $ \eta $, convergence threshold $ \epsilon $, and maximum available memory $ M_{\text{max}} $

\While{$ |L(B) - L(B_{\text{prev}})| > \epsilon $}
    \State $\nabla_B L(B) = \frac{\partial L(B)}{\partial B}$
    \State $ B \gets B - \eta \cdot \nabla_B L(B) $
    \If{$ M(B) > M_{\text{max}} \ \text{and} \ L(B) > T_{\text{real-time}} $}
        \State $ B \gets B / 2 $
    \EndIf

    \If{$ \text{sparsity}(T) > \text{threshold} $}
        \State $ B \gets \min(2B, M_{\text{max}}) $
    \ElsIf{$ \text{computational\_intensity}(T) > \text{threshold} $}
        \State $ B \gets B / 2 $
    \EndIf

    \State $ L(B_{\text{prev}}) \gets L(B) $
\EndWhile
\end{algorithmic}
\end{algorithm}

\begin{table*}[t]
\vspace{-0.3cm}
\centering
\footnotesize
\begin{tabular}{lccccccc} \hline
\textbf{Edge GPU}  & \textbf{AI Performance}  & \textbf{DRAM}  & \textbf{CPU} & \textbf{GPU}  & \textbf{Power} \\ \hline
Jetson Orin Nano        & 67TOPS (INT8)        & 8GB 102GB/s            & 6$\times$ARM Cortex-A78AE v8.2@1.7GHz   &  1024$\times$Ampere@1GHz    & 7-25W   \\
Jetson AGX Orin         & 275TOPS (INT8)       & 64GB 204.8GB/s           & 12$\times$ARM Cortex-A78AE v8.2@2.2GHz  &  2048$\times$Ampere@1.3GHz     & 15-60W  \\
\hline
\end{tabular}
\caption{Configurations of edge devices used in the evaluation.}
\label{tbl:device}
\end{table*}

\section{Evaluation} \label{evaluation}

\subsection{Prototype} 
\label{sec:implementation}
We implemented SparOA in 3,200 lines of Python. SparOA supports common DNN operators including convolution, fully connected, activation, normalization, pooling, and attention. The RL-based operator scheduler in Algorithm \ref{alg:sac_hybrid_inference} uses SAC from \texttt{stable baselines3}~\cite{raffin2021stable}. Training data is collected from runtime characteristics (sparsity, computational intensity, input/output size, and hardware utilization).

For the threshold predictor, we employ a Transformer-LSTM architecture with 128-dimensional hidden layers and 4 attention heads. The predictor takes sparsity, computational intensity, channels, height, and width as inputs, trained with Adam optimizer (learning rate $10^{-4}$) for 100 epochs. We use 80\% of samples for training and 20\% for testing.
\begin{table}[t]
\vspace{-0.3cm}
\centering
\footnotesize
\begin{tabular}{lccccc} \hline
\textbf{Model}  & \textbf{Parameters}  &\textbf{\makecell[c]{Computational \\ Intensity}}  & \textbf{\makecell[c]{Number of \\ Operators}} \\ \hline
ResNet-18          & 11.7M      & 1.8GFLOPs      & 53   \\
MobileNet-v3-small & 3.5M       & 0.3GFLOPs      & 112   \\
MobileNet-v2       & 2.5M       & 0.05GFLOPs     & 121   \\
ViT-B16            & 86M        & 17.6GFLOPs     & 65   \\
Swin Transformer   & 28M        & 4.5GFLOPs      & 125   \\
\hline
\end{tabular}
\caption{Configurations of DNN models.}
\label{tbl:model}
\end{table}

\subsection{Experiment Setup}
\label{sec:setup}

We execute experiments  on two edge devices configurations as detailed in Table~\ref{tbl:device}, representing both high-end and low-end hardware scenarios.

\fakepar{Models} We select five DNN models that are widely deployed in real applications, including ResNet-18~\cite{he2016deep}, MobileNet-v3-small~\cite{howard2019searching}, MobileNet-v2~\cite{sandler2018mobilenetv2},
ViT-B16~\cite{dosovitskiy2020image} and Swin Transformer~\cite{liu2021swin}. 
Table~\ref{tbl:model} lists the parameters, computational intensity, and number of operators.
We run each DNN model 10 times and compute average results.

\fakepar{Datasets} 
The input data are derived from ImageNet-2012~\cite{krizhevsky2017imagenet} and MS COCO-2014~\cite{lin2014microsoft}. 
The ImageNet-2012 dataset is used to evaluate the inference performance of CNN-based models, such as ResNet-18 and MobileNet. The MS COCO-2014 dataset is suitable for testing the performance of attention-based DNN models, such as Vision Transformer and Swin Transformer. The two datasets comprehensively evaluate the hybrid inference efficiency of SparOA under different architectures and input characteristics.

\fakepar{Baselines} 
We compare SparOA with:
\begin{itemize}
    \item CPU-Only: the whole model is executed on the CPU.
    \item GPU-Only (PyTorch)~\cite{paszke2019pytorch}: a deep learning framework with dynamic graph. PyTorch dispatches operators one by one sequentially.
    \item TensorFlow~\cite{abadi2016tensorflow}: a deep learning framework with static graphs. The scheduler in TensorFlow also dispatches operators sequentially.
    \item TensorRT~\cite{NVIDIATensorRT}: schedules the computation graph after the automatic fine-tuning of the kernel to the multi-stream in the GPU to execution in parallel.
    \item TVM~\cite{chen2018tvm}: a machine learning compiler framework that utilizes AutoTVM~\cite{chen2018learning} and AutoScheduler~\cite{zheng2020ansor} to generate efficient schedules.
    \item IOS~\cite{ding2021ios}: an inter-operator scheduler, which utilizes operator fusion and inter-operator parallelism to accelerate inference.
    \item POS~\cite{zhang2023pos}: a learning-based operator scheduler, which utilizes operator fusion, subgraph reuse, inter- and intra-operator parallelism to accelerate inference.
    \item CoDL~~\cite{jia2022codl}: a state-of-the-art CPU-GPU collaborative inference framework, accelerates inference through hybrid-type-friendly data sharing.
    \item SparOA w/o RL: a variant of SparOA without operator scheduling strategy.
    \item SparOA with Greedy: a variant of SparOA based on a greedy strategy for operator scheduling.
    \item SparOA with DP: a variant of SparOA based on dynamic programming for operator scheduling.
\end{itemize}

\begin{figure*}[tb]
\vspace{-0.2cm}
\centering
\subfigure[NVIDIA Jetson Orin Nano]{
\begin{minipage}[b]{\linewidth}
\includegraphics[width=\linewidth]{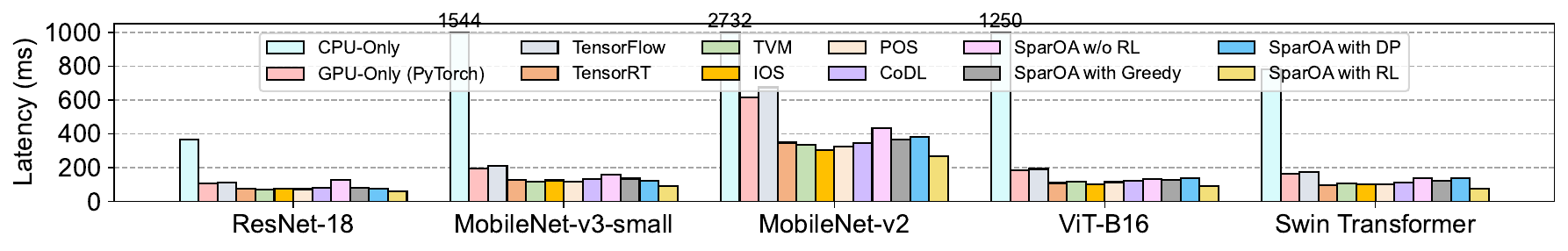}
\end{minipage}}
\subfigure[NVIDIA Jetson AGX Orin]{
\begin{minipage}[b]{\linewidth}
\includegraphics[width=\linewidth]{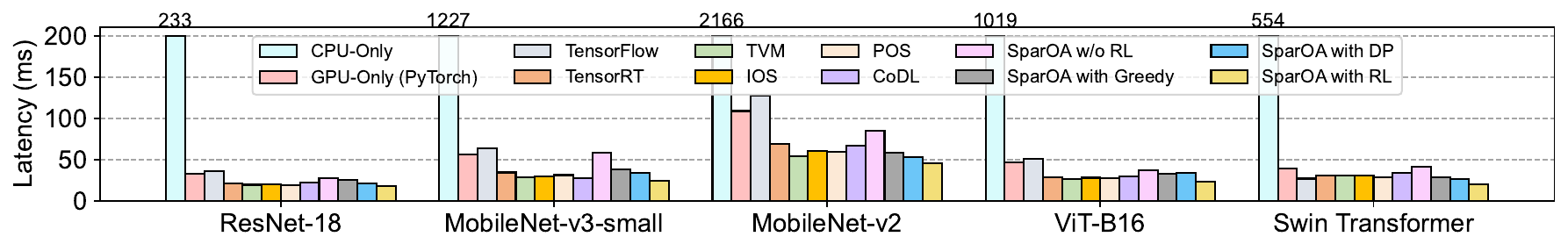}
\end{minipage}}
\vspace{-0.5cm}
\caption{Inference latency comparison of SparOA against the baselines, depending on target device.}
\label{fig:latency}
\vspace{-0.1cm}
\end{figure*}

\subsection{Overall Performance}\label{Overall}
\label{sec:overallPerformance}
Figure~\ref{fig:latency} reports the comparison of inference latency. 
For AGX Orin, SparOA achieves up to 50.7$\times$ speedup over baselines. The most significant improvement is against CPU-Only (e.g., 50.7$\times$ on MobileNet-v3). Against DNN compilation frameworks (e.g., TensorRT, TVM, IOS, POS) and the SOTA co-execution baseline (e.g., CoDL), SparOA achieves $1.22\times$$\sim$$1.31\times$ speedup. 
These SOTA baselines generate a fixed execution plan. In contrast, the SAC-based scheduler in SparOA can dynamically adapt to real-time hardware conditions, resulting in lower latency.

Futhermore, SAC-based RL scheduling further improves performance by $1.17\times$$\sim$$1.42\times$ over non-RL variants (Greedy, DP). For compute-intensive models (ViT-B16, Swin Transformer), SparOA reduces latency by $1.13\times$$\sim$$1.31\times$ over compilation frameworks by dynamically offloading high-computational intensity operators to GPU while executing sparse activation layers on CPU.
For Orin Nano, SparOA achieves $1.24\times$$\sim$$11.43\times$ acceleration, demonstrating its scalability across resource-constrained environments.

\begin{figure}[tb]
\vspace{-0.2cm}
\centering
\subfigure[NVIDIA Jetson Orin Nano]{
\begin{minipage}[b]{\linewidth}
\includegraphics[width=\linewidth]{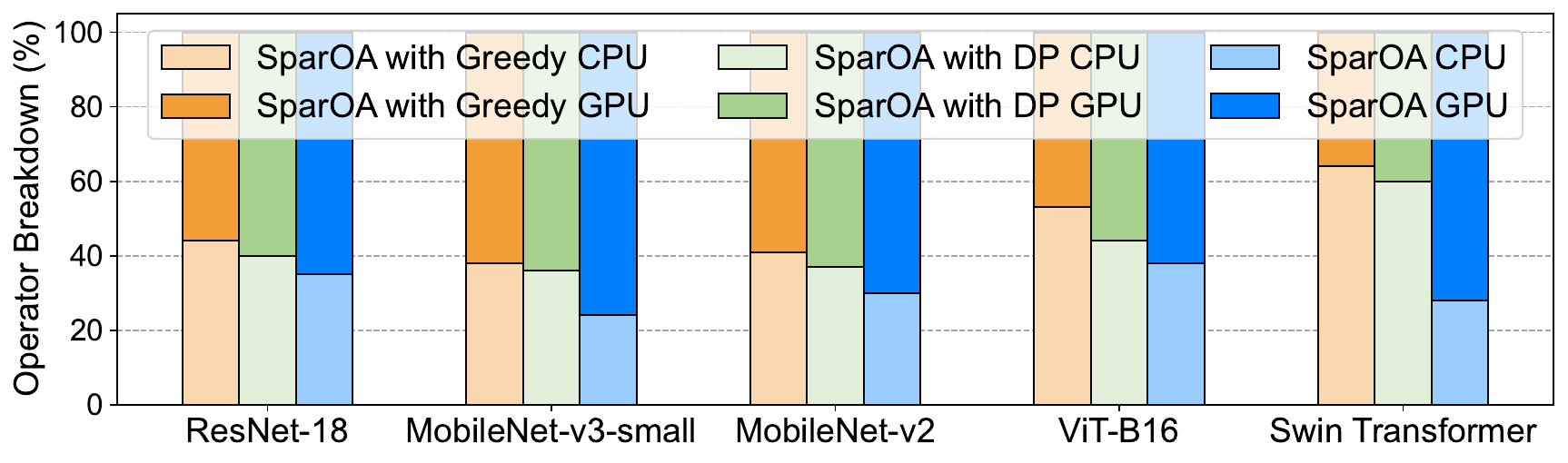}
\end{minipage}}
\subfigure[NVIDIA Jetson AGX Orin]{
\begin{minipage}[b]{\linewidth}
\includegraphics[width=\linewidth]{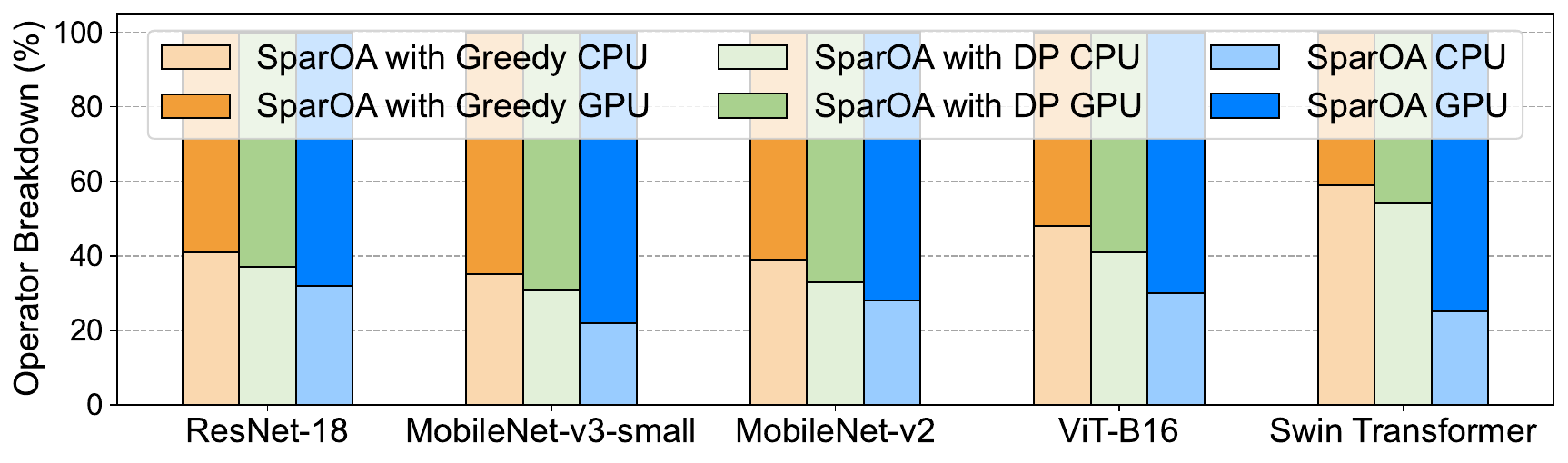}
\end{minipage}}
\vspace{-0.5cm}
\caption{Operator distribution on CPU and GPU during inference.}
\vspace{-0.5cm}
\label{fig:distribution}
\end{figure}

\begin{figure}[tb]
\vspace{-0.2cm}
\centering
\subfigure[NVIDIA Jetson Orin Nano]{
\begin{minipage}[b]{\linewidth}
\includegraphics[width=\linewidth]{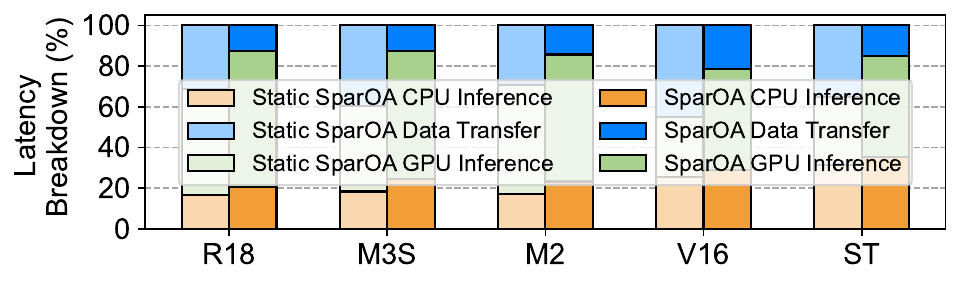}
\end{minipage}}
\subfigure[NVIDIA Jetson AGX Orin]{
\begin{minipage}[b]{\linewidth}
\includegraphics[width=\linewidth]{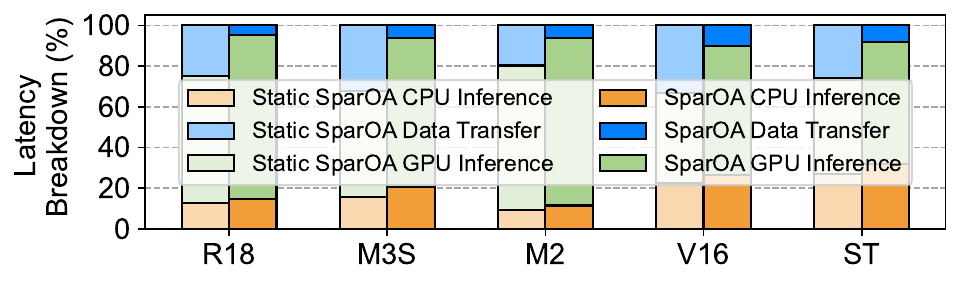}
\end{minipage}}
\vspace{-0.3cm}
\caption{Latency breakdown for static SparOA (i.e., without RL) and SparOA during inference.}
\vspace{-0.2cm}
\label{fig:transfer}
\end{figure}

As shown in Figure~\ref{fig:distribution}, SparOA's RL scheduler dynamically adjusts operator allocation based on real-time hardware states, increasing GPU operator load share to 72.6\% compared to 55.6\% (Greedy) and 60.8\% (DP). 
In Figure~\ref{fig:transfer}, this adaptive scheduling significantly reduces data transfer latency by $14.1\%$$\sim$$20.8\%$ compared to static scheduling (i.e., SparOA without RL), while dynamic batching optimizes batch sizes for different operator types to maximize resource utilization.

\begin{figure}[tb]
\vspace{-0.2cm}
\large
\centerline{\includegraphics[width=0.9\linewidth]{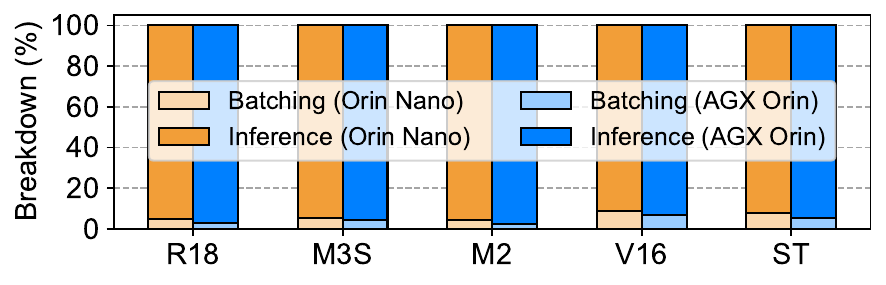}}
\vspace{-0.3cm}
\caption{End-to-end batching overhead of SparOA. The Y-axis displays the percentage breakdown between batching overhead and inference time.}
\vspace{-0.2cm}
\label{fig:batching}
\end{figure}

\subsection{Threshold Predictor}
We test the threshold predictor's performance through rigorous benchmarking against linear regression (LR) and CNN-based predictors, as summarized in Table 3. Our Transformer-LSTM threshold predictor achieves 92.3\% accuracy on Sparsity and 90.6\% on Computational Intensity ($\pm10\%$ tolerance), significantly outperforming CNN (36.2\% and 38.5\%) and linear models (23.7\% and 20.4\%). It combines Transformer and LSTM to capture both global dependencies and temporal correlations across operators. For Swin Transformer's attention layers, it reduces threshold estimation errors from 13.8\% to 2.1\%. Despite this high accuracy, our predictor remains lightweight ($\sim$4MB), enabling effective deployment on resource-constrained devices.

\begin{table}[tb]
  \centering
  \vspace{-0.1cm}
  \begin{tabular}{ccccc}
    \toprule
    \multirow{2}{*}{\textbf{Predictor}} & \multicolumn{2}{c}{\textbf{Metric}} & \multirow{2}{*}{\textbf{Model Size}} \\
    \cmidrule(lr){2-3}
                        & Sparsity &\makecell[c]{Computational \\ Intensity} & \\
    \midrule
    LR   & 23.7\%  & 20.4\%      & $\sim$7B     \\
    CNN          & 36.2\%  & 38.5\%      & $\sim$0.5MB \\
    \textbf{Ours} & \textbf{92.3\%} & \textbf{90.6\%}  & \textbf{$\sim$4MB}   \\
    \bottomrule
  \end{tabular}
  \caption{$\pm$10\% prediction accuracy and model size of the threshold predictors.}
  \label{tab:predictors}
\end{table}

\subsection{Hybrid Inference Engine}  
Figure~\ref{fig:batching} illustrates the end-to-end batching overhead of SparOA on Orin Nano and AGX Orin. 
Using CUDA streams, it reduces switching overhead by 52\% and achieves 78\% overlap between data transfer and computation. The gradient-based batching algorithm dynamically adjusts batch sizes (1-512), reducing batching overhead to $2.3\%$$\sim$$8.6\%$ compared to $15.4\%$$\sim$$28.7\%$ in static frameworks. For ViT-B16, it selects batch size 32 for attention operators and 8 for activation operators to balance GPU occupancy and memory usage.

The engine employs a weighted averaging strategy (e.g., $P = 0.7 \cdot P_{\text{gpu}} + 0.3 \cdot P_{\text{cpu}}$) that reduces aggregation errors by 14\%. SparOA achieves 89\% GPU utilization for ViT-B16 (vs. 63\% in TensorRT) and reduces memory usage by 37\% for MobileNet-v2, enabling efficient deployment on resource-constrained edge devices.

\subsection{Ablation Study}
\begin{figure}[tb]
\vspace{0pt}
\setlength{\abovecaptionskip}{0pt}
\setlength{\belowcaptionskip}{0pt}
\centering
\subfigure[NVIDIA Jetson Orin Nano]{
\begin{minipage}[b]{0.4825\linewidth}
\includegraphics[width=1\linewidth]{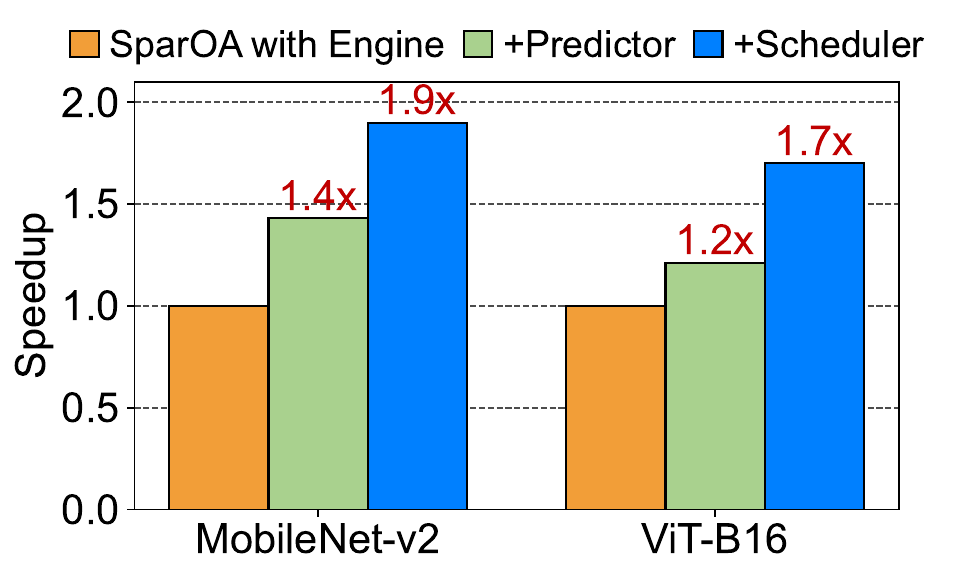}
\end{minipage}}
\subfigure[NVIDIA Jetson AGX Orin]{
\begin{minipage}[b]{0.4825\linewidth}
\includegraphics[width=1\linewidth]{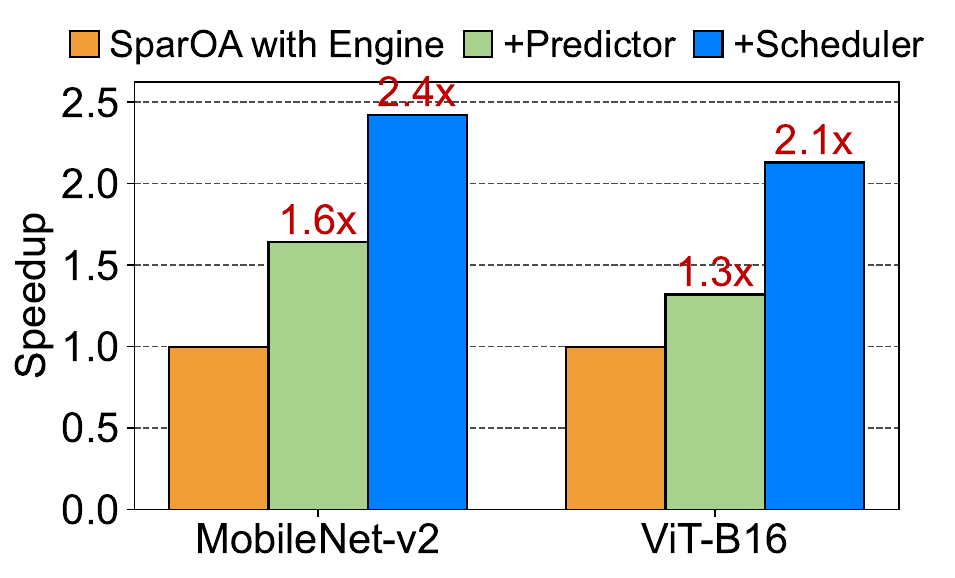}
\end{minipage}}
\caption{Inference performance breakdown.}
\vspace{-0.3cm}
\label{fig:ablation}
\end{figure}

Figure~\ref{fig:ablation} quantifies the contribution of each SparOA component. We evaluate MobileNet-v2 (CNN) and ViT-B16 (Transformer) on both edge devices.
The baseline hybrid engine (SparOA w/o Predictor/Scheduler) is normalized.

Threshold Predictor (+Predictor) dynamically adjusts sparsity/computational intensity thresholds, reducing redundant computations (e.g., pruning low-contribution kernels in 
Mobile\-Net-v2). Gains are higher for 
MobileNet-v2 ($1.4\times$$\sim$$1.6\times$ 
speed\-up) than ViT-B16 due to the latter’s sparse attention variability.
Learning-based Scheduler (+Scheduler) prioritizes GPU offloading for compute-intensive operators (e.g., depthwise convolutions, matrix multiplications), further improving speed\-ups (MobileNet-v2: $1.9\times$$\sim$$2.4\times$; ViT-B16: $1.7\times$$\sim$$2.1\times$). Hardware constraints limit gains on Orin Nano, where memory bottlenecks restrict batch optimization.

Overall, MobileNet-v2 benefits most from sparsity optimization and GPU offloading.
ViT-B16 relies on scheduler-driven resource allocation for compute/memory balancing.
High-end devices unlock parallelism, and low-end devices face memory limits.

\subsection{Convergence Time}
\label{sec:Convergence}

\begin{figure}[tb]
\large
\centerline{\includegraphics[width=0.85\linewidth]{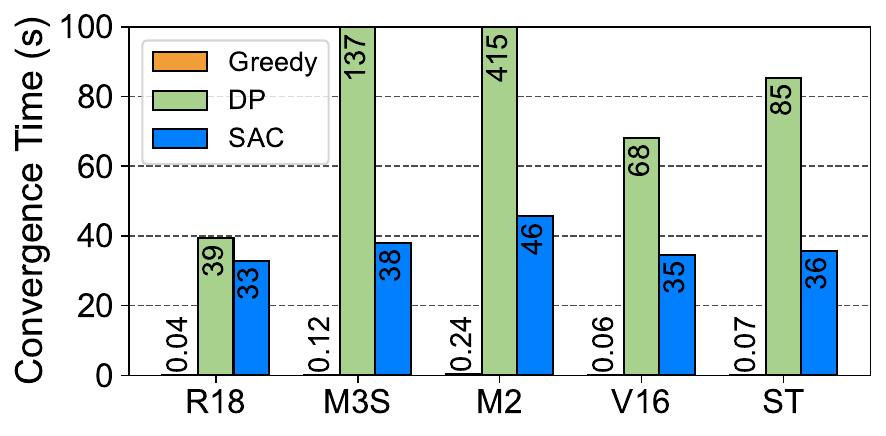}}
\vspace{-0.5cm}
\caption{Convergence time of different scheduling algorithms on NVIDIA Jetson AGX Orin.}
\vspace{-0.2cm}
\label{fig:convergence}
\end{figure}

Figure~\ref{fig:convergence} compares convergence times of scheduling algorithms. Greedy converges rapidly ($0.04$$\sim$$0.24s$) but ignores hardware states, resulting in 22\% higher latency than SAC. DP requires excessive time ($39$$\sim$$415s$) due to exhaustive search, yet yields suboptimal strategies (e.g., 63ms vs SAC's 48ms for MobileNet-v2).

SAC achieves optimal performance with $33$$\sim$$46s$ convergence time. Though slower to converge than Greedy, its dynamic adaptation to hardware states delivers superior runtime efficiency. Notably, SAC demonstrates sublinear convergence growth with model complexity, enabled by its parallel policy evaluation mechanism. This trade-off proves valuable as the convergence overhead is amortized over numerous inference executions.

\subsection{System Overhead}
\label{sec:overhead}
\subsubsection{Power/Energy Consumption.}
\begin{figure}[tbp]
\vspace{-0.2cm}
\centering
\subfigure[Power consumption per inference]{
\vspace{-0.5cm}
\begin{minipage}[b]{\linewidth}
\includegraphics[width=\linewidth]{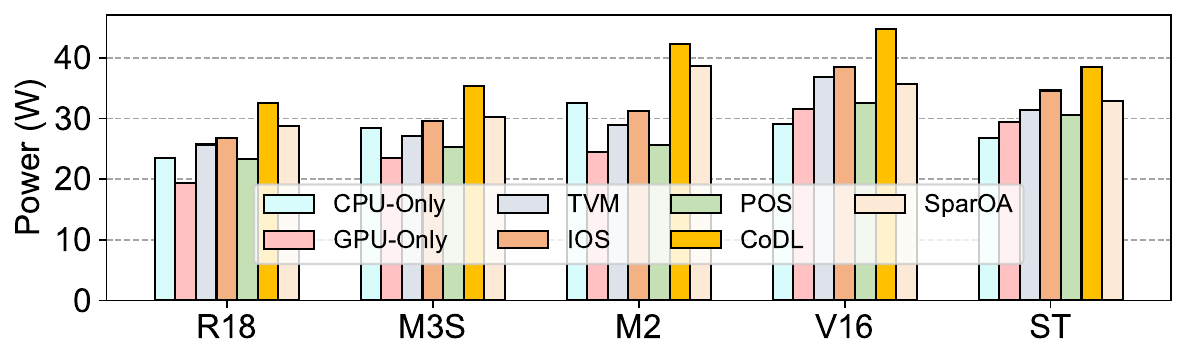}
\end{minipage}}
\subfigure[Energy consumption per inference]{
\begin{minipage}[b]{\linewidth}
\includegraphics[width=\linewidth]{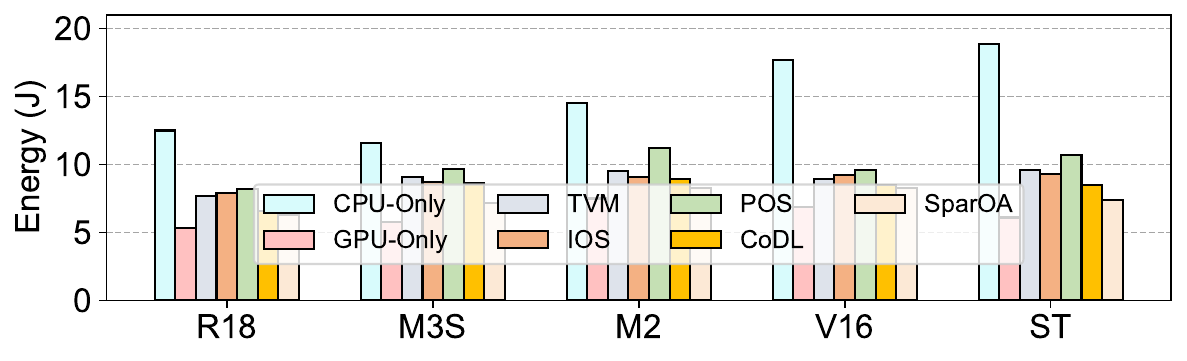}
\end{minipage}}
\vspace{-0.3cm}
\caption{Power/energy consumption measurements on NVIDIA Jetson AGX Orin.}
\label{fig:power}
\vspace{-0.3cm}
\end{figure}

As shown in Figure~\ref{fig:power}(a), SparOA’s power consumption is higher than single-processor baselines. This is a necessary trade-off for activating both CPU and GPU resources. Also, SparOA consumes slightly more power than compiler frameworks such as TVM (e.g., 34\% higher) and IOS (e.g., 24\% higher). Notably, SparOA’s power draw is consistently more efficient than CoDL.
In Figure~\ref{fig:power}(b), SparOA achieves the lowest energy-per-inference. This superior energy efficiency, resulting from a controlled power trade-off, leads to a significant reduction in latency.

\subsubsection{Memory Usage.}
\begin{figure}[tbp]
\vspace{-0.1cm}
\large
\centerline{\includegraphics[width=1\linewidth]{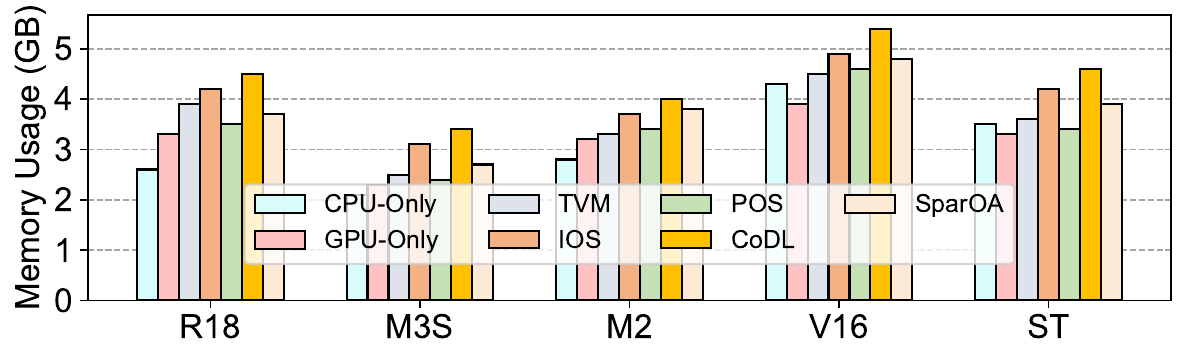}}
\vspace{-0.3cm}
\caption{Memory usage measurements, using the NVIDIA Jetson AGX Orin as the test device.}
\vspace{-0.3cm}
\label{fig:memory}
\end{figure}

In Figure~\ref{fig:memory}, SparOA employs a sharded storage strategy for co-execution (i.e., CPU stores sparse attention heads while GPU stores dense weights). This results in an average of 23.1\% increase in memory overhead compared to GPU-Only. While SparOA requires more memory than single-processor methods, its memory footprint is comparable to other baselines (e.g., IOS and POS), and is lower than CoDL. This modest memory overhead is an explicit trade-off for enabling hybrid scheduling and is justified by the substantial improvements in inference latency.

\section{Related Work}\label{Related}
\fakepar{Optimization of DNN Inference}
Model adaptations, including model compression~\cite{han2015deep}, pruning,  and quantization~\cite{he2018amc,wang2019haq,lin2024awq,lin2024qserve}, reduce computational complexity but often sacrifice accuracy and fail to exploit heterogeneous hardware. DNN compilers~\cite{ma2020rammer,zhu2022roller,shi2023welder,zhang2023cocktailer} like TVM~\cite{chen2018tvm}, Ansor~\cite{zheng2020ansor}, TensorRT~\cite{NVIDIATensorRT}, IOS~\cite{ding2021ios}, and POS~\cite{zhang2023pos} generate optimized, hardware-specific code, but typically target a single processor. 
Other works address orthogonal challenges: AdaptiveNet~\cite{wen2023adaptivenet} tackles device diversity by selecting an optimal subnet from a supernet post-deployment, while FlexNN~\cite{li2024flexnn} manages memory constraints by slicing models that exceed RAM. SparOA is distinct from these approaches, we optimize a single, fixed model assumed to be in memory, focusing on hybrid operator scheduling across the CPU and GPU, rather than model selection or I/O management.

\fakepar{Co-execution of heterogeneous processors}
Edge devices feature diverse processors~\cite{jia2022codl}, which several frameworks aim to exploit. CoDL~\cite{jia2022codl} optimizes inference latency through dynamic operator scheduling based on processor affinity. Other systems partition tasks for multi-DNN inference (e.g., Band~\cite{jeong2022band} and BlastNet~\cite{ling2022blastnet}), automatically branch models (NN-Stretch~\cite{wei2023nn}), or split workloads for specific tasks and hardware~\cite{huang2024perceptual,wang2024intelligent,bao2024couple}. However, these methods, including CoDL, primarily rely on static processor characteristics for scheduling. They largely overlook how intrinsic model properties, such as dynamic sparsity and computational intensity, affect performance. SparOA addresses this limitation by incorporating both metrics to achieve a more fine-grained operator-processor matching.

\fakepar{DNN performance prediction}
Existing performance predictors range from kernel-level mathematical models~\cite{li2023predicting,li2023path} (e.g., nn-Meter~\cite{zhang2021nn}, nnPerf~\cite{chu2023nnperf}) to ML-based predictors that learn from historical data (e.g., MAPLE-Edge~\cite{nair2022maple}, LitePred~\cite{feng2024litepred}). These methods often ignore heterogeneous processor characteristics or require extensive data, focusing on predicting latency for a single platform. SparOA's threshold predictor is fundamentally different. It is not a general-purpose latency predictor; rather, it is a specialized, lightweight model designed to capture the non-linear relationship between sparsity and computational intensity to accurately determine the optimal scheduling thresholds for hybrid CPU-GPU execution.

\section{Discussion} \label{sec:Discussion}
\textbf{Hardware Heterogeneity.}
Extending SparOA to support more heterogeneous AI accelerators~\cite{jeong2022band} (e.g., NPUs, TPUs) presents significant challenges. The operator scheduler would need to learn more complex allocation policies across multiple hardware types with vastly different characteristics. One promising approach is that SparOA integrates hierarchical reinforcement learning~\cite{nachum2018data,qin2024earnhft} that first determines broad hardware categories before making finer-grained decisions within the accelerator cluster. This approach would maintain scheduling efficiency while expanding hardware support.

\fakepar{Training Overhead and Model Adaptation}
When DNN architectures change significantly, the predictor may require complete retraining. To address this, SparOA can be combined with transfer learning~\cite{zhou2024dynamic,xin2024vmt}, which enables the predictor to quickly adapt to new models with minimal additional training by leveraging knowledge from previously seen architectures. Additionally, meta-learning~\cite{wei2024free,feng2024bioactivity} could further enable SparOA to generalize across model families, reducing the training overhead for novel architectures.

\fakepar{Multi-Task Inference}
Edge devices typically involve multiple concurrent tasks with varying service-level objectives (SLOs)~\cite{kim2023dream}. In multi-DNN inference~\cite{liu2022veltair,zhang2023pos}, the scheduler in SparOA cannot prioritize critical tasks. To this end, we will extend SparOA with a multi-level scheduler that incorporates task priority metrics and SLOs. Meanwhile, combining SparOA with techniques like knowledge distillation~\cite{puy2024three} could create lightweight versions of DNNs for non-critical tasks, while reserving full resources for high-priority models. Furthermore, integrating with edge orchestration systems like Kubernetes (k8s)~\cite{burns2016borg} could enable coordinated resource management across multiple applications running on the same edge device.

\section{Conclusion}\label{conclusion}
This work presents SparOA, a framework that redefines hybrid inference on edge devices by treating sparsity and computational intensity as orthogonal, joint scheduling metrics. SparOA integrates a lightweight predictor with a dynamic RL-based scheduler, enabling it to adapt to real-time hardware fluctuations.
Extensive evaluations on NVIDIA Jetson platforms demonstrate that SparOA outperforms SOTA compilers and co-execution baselines in terms of inference latency and energy efficiency.


\bibliographystyle{ACM-Reference-Format}
\bibliography{ref}

\end{document}